# DSH-RPL: A Method based on Encryption and Node Rating for Securing the RPL protocol Communications in the IoT Ecosystem

Mina Zaminkar[1] . Fateme Sarkohaki[2] . Reza Fotohi[3]

[1]Department of Computer Engineering, Faculty of Engineering, Shahid Ashrafi Esfahani University, Isfahan, Iran

[2]School of Computer Engineering, Iran University of Science & Technology, Tehran, Iran

[3]Faculty of Computer Science and Engineering, Shahid Beheshti University, Tehran, 1983969411, Iran

**Correspondence**
Faculty of Computer Science and Engineering, Shahid Beheshti University, Tehran, 1983969411, Iran

Email: Fotohi.reza@gmail.com;
R_fotohi@sbu.ac.ir

**Abstract**

Internet of Things (IoT) provides the possibility for milliards of devices throughout the world to communicate with each other, and data is collected autonomously. The big data generated by the devices should be managed securely. Due to security challenges, like malicious nodes, many approaches cannot respond to these concerns. In this paper, a robust hybrid method, including encryption, is used as an efficient approach for resolving the RPL protocol concerns so that the devices are connected securely. Therefore, the proposed DSH-RPL method for securing the RPL protocol comprises the four following phases: The first phase creates a reliable RPL. The second phase detects the sinkhole attack. The third phase quarantines the detected malicious node, and the fourth phase transmits data through encryption. The simulation results show that the DSH-RPL reduces the false-positive rate more than 18.2% and 23.1%, and reduces the false-negative rate more than 16.1% and 22.78%, it also increases the packet delivery rate more than 19.68% and 25.32% and increases the detection rate more than 26% and 31% compared to SecTrust-RPL and IBOOS-RPL.

## 1 | INTRODUCTION

IoT, which is known as the fourth industrial revolution, connects various physical devices, smart counters, wireless thermometers and smart vehicles, and has developed various technologies in the human society using new and modern methods, in which each object can receive and transmit data via communication networks, including the Internet [1]. Security in information exchange is a necessity that has attracted attention since ancient times. Considering the emergence of wireless technology and its widespread use due to easy and fast implementation and low cost compared to the implementation of the network's physical platform and security shortcomings in this technology and possibility of malicious attacks and phishing, necessitates presenting approaches to handle these attacks [2]. Therefore, considering the nature of the IoT ecosystem's wireless technology, which has many vulnerabilities, considering the weak points and trying to resolve them might speed up using this ecosystem. Security in IoT, like other networks, is based on confidentiality and trust. Thus, attack detection systems are one of the primary defense methods against attacks in IoT. One of the dangerous attacks in IoT is the sinkhole attack. The Sinkhole is one of the most dangerous attacks that operate in the network layer of the protocol stack. In this attack, the malicious node tries to attract the network traffic by broadcasting fraud routing information and then it does not forward the packets.

Therefore, this study's main contribution in resolving the above problems and making communications among IoT devices secure is to introduce a secure method based on ranking and encryption to solve the mentioned concerns. Thus, in this paper, a method is presented that solves the above problem in the following four steps:

Step 1: The first phase creates a reliable RPL.

Step 2: The second phase detects the sinkhole attack.

Step 3: The third phase quarantines the detected malicious node,

Step 4: and the fourth phase transmits data through encryption.

The remainder of the paper is structured as follows. Section II reviews the research background on securing communication between devices. Section III discusses the proposed DSH-RPL in detail. Section IV deals with the simulation and performance evaluation of the proposed method. Finally, Section V concludes the paper.

## 2 | SINKHOLE ATTACKS AND DETECTION APPROACHES
This section explains the concepts of sinkhole attack and detection approaches.

### 2.1 | SINKHOLE ATTACK
The sinkhole is one of the most dangerous attacks in IoT networks, which takes place in the network layer of the protocol stack and prevents the packets from reaching the destination. In a sinkhole attack, the intruders aim to absorb the total traffic by a node originating from one region. Since the nodes are close to or along the route along which the packets travel, the chance that the packets can be accessed is very high. Sinkhole attacks can be a means for other attacks (selective forwarding). Sinkhole attacks are usually performed by a node, which is more attractive than other nodes concerning the routing algorithm. One motivation for the sinkhole attack might be the selective forwarding attack. Since the total traffic streams from one node, an intruder can omit or manipulate the packets originating from all nodes of a specific region. Figure 1 shows a sinkhole attack in an IoT ecosystem.

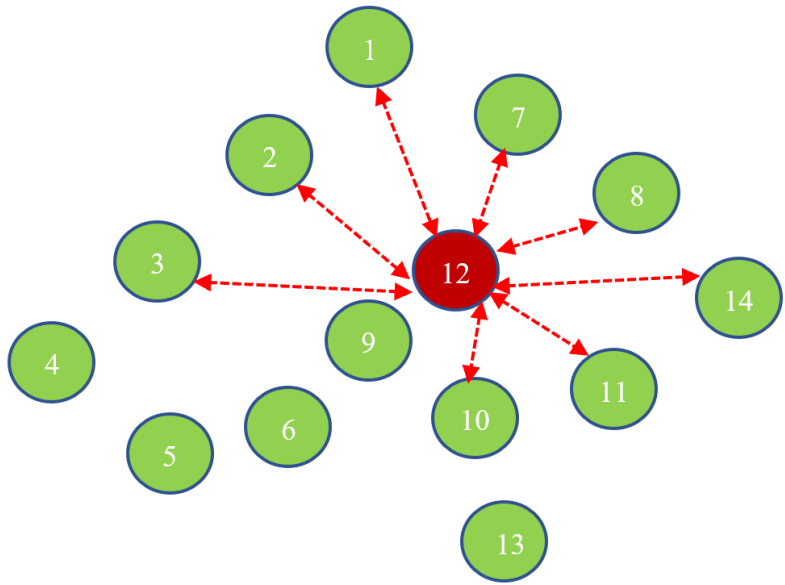

**FIGURE 1** Sinkhole attack in IoT.

## 2.2 | Detection approaches

A large number of studies have been conducted on securing communication between IoT devices. According to the subject matter of the present research, several prominent studies performed in this domain will be reviewed and evaluated.

In [3], a method has been presented for detecting the malicious nodes in IoT. This hybrid intrusion detection model is used to detect the malicious operations of "selective forwarding" and "sinkhole attack" based on the 6LOWPAN protocol.

In [4], the proposed defensive approach includes the following:
1. The average energy consumption and memory usage (RAM) are saved significantly.
2. The proposed method reduces the number of IDSs without threatening their efficiency.
3. The proposed method is based on the 6LoWPAN and employs the local knowledge, which makes it scalable and distributed.
4. The proposed method maximizes the true positive rate and the true negative rate.

In [5], the Blockchain Lightweight Scalable (LBS) has been proposed, which has been optimized for IoT requirements and in the meantime, ensures end-to-end security. The proposed blockchain has established an overlay network on which the devices that jointly control blockchain management along with the upper resource may achieve a distributed mechanism. The overlay network has been organized to reduce principal costs as certain clusters and cluster-heads are responsible for public chain management. This method utilizes a distributed trust mechanism exploited by cluster-heads to diminish data processing overload gradually for verifying new blocks.

A secure method has been presented in [6] to develop an efficient IDS for the IoT devices, including the following structure: The IDS is comprised of three sections: observation, analysis, and alarm. The proposed IDS is a detection method based on the anomaly, suitable for use in IoT. In this method, if an intrusion is detected, the IDS informs the users before the adversary attacks. Also, the proposed method can detect intrusion through observing the network and the connected devices. The proposed IDS detects the routing attacks that have not been detected before. Considering the method presented in this paper for preventing the active sinkhole attack, VoIP devices do not communicate directly so that fewer resources are consumed. The efficiency of the proposed method demonstrates that the computational overhead is reduced for all networks, reasonably. Also, its detection accuracy for routing attacks is high.

In [7], a distributed mechanism based on access control and authentication has been presented for IoT devices, which can be applied for different scenarios under certain conditions. It is based on the notion of "blockchain" and "fog computing technology". Taking advantage of fog computing characteristics and the distributed nature of blockchain, a novel delay-sensitive mechanism with blockchain capabilities was proposed for IoT devices. The main contribution of this paper is the introduction of a decentralized mechanism that allows for the validation and access control to create a controlled and secure environment in which the devices communicate with secure information exchange. This communication can be established devices belonging to various systems or between identical devices.

The method presented in [1] is based on a trust-aware RPL that can track and isolate the routing attacks with acceptable performance. The proposed method is called SecTrust-RPL, which is a routing protocol based on trust RPL, formulated based on the previously proposed framework of the authors, SecTrust.

The innovations of this paper are as follows:
1. This method validates the nodes.
2. The proposed framework is based on the RPL routing protocol, which presents a new trust-aware RPL protocol.
3. The effectiveness of the SecTrust-RPL is compared with the standard RPL protocol through simulation under sinkhole attacks.
4. Also, a real experiment is presented on the physical test platform for SecTrust-RPL to validate the simulation results.

Authors of [8] have used the fuzzy logic to detect the malicious nodes in IoT and restricted their untrustworthy performance for network nodes. This method is based on clusters in the IoT network, fuzzy security protocol, and trust management via communication and secure message exchange among the IoT nodes. The proposed protocol employs a message system with serial communications for secure message encryption that allows the nodes to move securely from one cluster to another.

In [9], the aim is to give a detailed explanation of the mechanism of smart contracts and blockchains, as well as to identify pros and cons, whose introduction would lead to a robust system, highlighting blockchain and IoT application systems together. The existent devices on IoT devices can serve as points of contact with the physical world. When combined, these devices can ensure accurate encryption and significant cost and time savings during data processing. The authors have also maintained that reading this article allows the reader to detect potentially new applications of their IoT-related activities and make rational decisions when incorporating blockchain technology into their projects.

Authors of [10] have presented a new method that decreases the sinkhole attacks and prevents saturation issues. Therefore, the key role of this paper is as follows:
1. SDELM model is used to provide IoT security and detect malicious attacks.
2. A new reduction algorithm is presented, which is classified as approximate algorithms and has not been presented in previous studies.

3. The newly proposed mechanism is tested in a real hardware network to evaluate its performance in practice.

4. A set of experiments is executed using the UNB-ISCX measure data, and the results are compared with the existing approaches.

In [11], the authors have presented an equitable access framework. It has been earmarked for an IoT system that lets a number of individuals and institutions to access each other's data. All members have been connected to the same blockchain network by which data access control has been performed. Every network member has a "purse" containing access information for the blockchain network members as well as all the keys to the information authorized to be observed. In such systems, if X requests a resource being processed by Y, the former must send its request along with all the corresponding keys to the blockchain network. Afterward, the blockchain network probes X's access permission to Y's resources. If X is permitted to receive the information, Y sends the data to that member.

In [12], a light penetration detection algorithm based on RPL protocol has been presented to defend Sybil and sinkhole attacks, which requires fewer computations, and its accuracy is high, which is essential in networks with limited resources. In the proposed method, all three types of Sybil attacks in stationary RPL and mobile RPL are studied, and a light penetration detection method is presented. The efficiency of the proposed algorithm is studied for all three types of Sybil attacks in terms of accuracy, sensitivity, and F-measure. Also, the authors have proposed a mathematical model inspired by nature for the Sybil attack in mobile RPL based on the artificial bee colony (ABC) model.

## 3 | PROPOSED DSH-RPL SCHEMA

The presence of malicious nodes has always been a concern in all networks, particularly the IoT ecosystem. Various methods have been proposed to protect these networks. In IoT, due to the presence of malicious nodes like the Sinkhole attack, the trust measure among the nodes participating in routing is a significant issue. The purpose of this study is to introduce a method for routing in a trustworthy RPL in IoT technology that detects malicious sinkhole nodes. The proposed method, called DSH-RPL[3], detects sinkhole attack in a trustworthy RPL. The proposed method is introduced in four phases. The first phase creates a trustworthy RPL, the second phase detects the sinkhole attacks, the third phase quarantines the detected malicious node, and finally, the fourth phase transmits data by encryption.

This paper is an improvement over a paper published by the authors of the same paper in [13], the differences between which are clarified as follows:
- In our previous work [13], only malicious nodes were discovered based on rankings, but in the new work, the following changes have been applied:
- The DODAG structure proposed in the new article is based on the Things reliability review, which is based on the criteria of energy, trust and honesty. This reliability and reliability review is an important and new topic, while in the previous article, the DODAG structure each of these steps was used without any review.

---
[3] Detection of sinkholes in RPL

- In the previous article, only the rank difference was used in the malicious node detection step, but in the present article, this rank difference was used to create the DODAG tree, based on reliability and PDR, to more accurately detect the attack.
- In the present paper, after removing the malicious node from the network to ensure data transmission, homomorphic encryption is used to send data, which did not exist in the previous paper.

### 3.1 | System Model of the Proposed Method

The proposed IoT network is comprised of N IoT nodes with limited sources $N = \{N_1, N_2, N_3, ...., N_n\}$, which can be sensors or mobile objects. These nodes are in the network area with homogeneous sources. However, the sources might be heterogeneous depending on their operation; that is, they might have different infrastructure hardware or software.

RPL creates a virtual DODAG on the network topology. For experimental purposes, the presence of sinkhole malicious nodes is also considered. The root node plays an essential role in creating and preserving the balance of the existing network. The root node is a border router (BR2) that takes information from the object and transmits it to the CLOUD for processing. As mentioned in most studies, it is assumed that the root node (BR) is trustworthy and cannot be considered a malicious node. Other nodes of the network might be malicious, which should be detected and omitted from the routing process. Figure 2, shows the general schematic of the system model of the proposed DSH-RPL method.

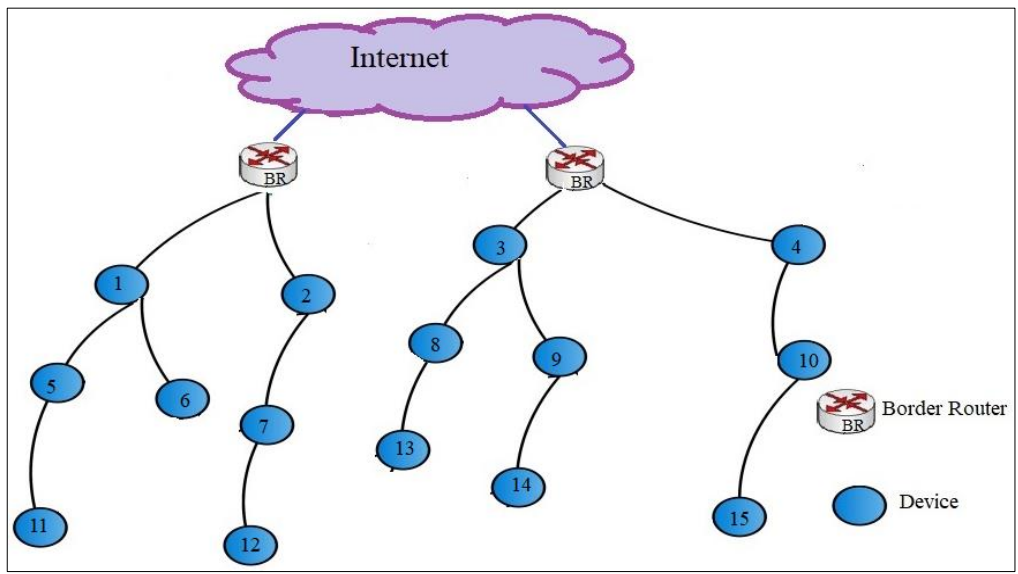

**FIGURE 2** System Model of the proposed DSH-RPL method

### 3.2 | Phase 1: Developing a Trustworthy DODAG

In the proposed method, the reliability among nodes is studied to develop the DODAG. To this end, several measures of each node, including energy, trust, and integrity, are investigated. These measures allow a node to communicate with trustworthy nodes only. However, as a result of this solution, the objects do not consume resources and cannot be assigned to IoT devices with limited energy resources, it is suggested that all computations are assigned to a TPM. The trustworthy platform module "TPM" is a small and low-cost security device aiming to execute the essential security services for a client machine. This hardware element keeps information like digital keys, certificates, and passwords securely. Such information is used to authenticate the platform. This process solves issues like signing messages, generating keys, and storage problems. Also, it can store the platform indices that help reliability [1].

To create a reliable DODAG, TPM is used to secure the control messages through encryption. Before exchanging control messages, a trustworthy relationship is created among nodes. The switching exchange is done by RSA keys. Also, it is suggested to use TPM as a cooperator processor on a node for secure and safe transmission of computations and storage operation. The proposed DSH-RPL method is based on two complementary points:

- It employs encryption-based methods and the authentication presented by TPM. These methods ensure trust among nodes for transmitting control messages.

- The node behavior is investigated to ensure that the nodes participating in the creation and maintenance of the DODAG topology are trustworthy.

### 3.2.1 | Reliability of the Nodes

To study network reliability and create a trustworthy DODAG, each node existing in any route of the network receives the reliability packet, REQP-R, generated by BR and inserts its information in the packet. Then, the node transmits the packet to its neighbors. Upon receiving this message from a node, the node's ID is recorded in the monitoring table, and its trust field is increased by one unit (the initial value of the trust field is zero). The node has recorded information of the REQP-R packet, including the node's ID and energy of the node, and increases its trust field. The Trust field indicates that the node is not selfish and transmits the packets to the other nodes. Therefore, the node that transmits the REQP-R packet to other nodes is trustworthy. Upon receiving REQP-R, each node checks if the packet is repetitious or not (if the ID of the node is listed at the origin route). If so, the packet is discarded, but the trust field of the transmitter node is updated. Otherwise, the node inserts its information in the REQP-R packet and transmits it to its neighbors. Then, it transmits the ACK packet back to the BR.

The trust field of a specific node is increased by the adjacent nodes whenever the node forwards the REQP_R packet, and its energy level is recorded in the monitoring table. The monitoring table is stored in the TPM of the node.

Each node should transmit its information and its monitoring table to the node transmitting the REQP_R, through transmitting the ACK packet. According to this information and the monitoring table information, a node's being veridical is specified, and its veracity filed in the monitoring table is updated. This field shows if a node is veridical about the information of its neighboring nodes or not, through receiving ACKs and the monitoring table of the node, and comparison with other nodes. Thus, each node can gather more knowledge about the behavior of other nodes. The format of REQP_R packet is shown in Figure 3.

| 1B | 1B | 2B |
|---|---|---|
| Type | code | Checksum |
| Node's ID ||| 
| Energy ||| 
| Source IP address ||| 
| Source sequence number ||| 

**FIGURE 3** Format of REQP_R packet

### 3.2.2 | Steps of Creating a Trustworthy DODAG

After creating the monitoring table, the BR creates the DODAG following the steps below:

**Step 1:** BR introduces itself as a floating root. Thus, it has no parents.

**Step 2:** (Calculating reliability): first, each node i evaluates the reliability of its neighboring node j at time t using Eq. (1) using three measures of energy, trust, and veracity of the monitoring table.

$$R_{ij}(t) = w_1 R_{ij}^{Energy}(t) + w_2 R_{ij}^{Trust}(t) + w_3 R_{ij}^{Veracity}(t) \quad (1)$$
$$w_1 + w_2 + w_3 = 1$$

In Eq. (1), $R_{ij}(t)$ is between 0 and 1 ($R_{ij}(t) \in [0,1]$). 1 indicates complete reliability, and 0 indicates non-reliability. $w_1$, $w_2$ and $w_3$ are the energy, trust, and veracity weights considered for reliability. These weights depend on the context and are defined by the network manager. Each element of the reliability values $x \in \{Energy, Veracity, Trust\}$ is evaluated considering the monitoring table and according to Eq. (2); where, $\Delta t$ is the reliability updating interval; $\alpha \in [0,1]$ indicates that evaluating reliability relies on direct or older observations. It was suggested that the nodes store their old observations in TPM.

$$R_{ij}^{x}(t) = \alpha R_{ij}^{x,direct}(t) + (1-\alpha) R_{ij}^{x}(t - \Delta t) \quad (2)$$

When node i receives DIO messages from its neighbors, it calculates the new reliability for each neighboring node j. The new reliability is the mean of the reliability values calculated using Eq. (1). Moreover, each node transmits the reliability to its neighbors via the DIO message. The obtained result shows the ultimate reliability for the neighboring node j, and it is used to select the set of parents.

$$R_j(Final) = \frac{\sum_{k=1}^{m} R_{kj}}{m} \tag{3}$$

In Eq. (3), $k \in \{i, (neighbors\ of\ i \cap neighbors\ of\ j)\}$ and m represents the number of the nodes from which the node i has received the reliability value of its neighboring node j.

Third, using Eq. (4), the node i calculates its new reliability, which is the mean value of all received reliability values calculated by the neighbors.

$$R_i = \frac{\sum_{k=1}^{n} R_{ki}}{n} \tag{4}$$

In Eq. (4), $k \in \{neighbors\ of\ i\}$ and n represents the number of neighboring nodes from which the node i has received its reliability value.

**Step 3 (Calculating the route cost based on reliability):** each node i calculates the route cost for each accessible neighbor j. This cost represents the cost of the route from node i to BR via node j regardless of the reliability measure. The route cost is $R_j(Final)$, which is calculated using Eq. (3). The route, including nodes with higher reliability values, is selected. However, the route, including trustless nodes, should be avoided. Therefore, the selected route might be the longest but the most secure one.

**Step 4 (Selecting parents):** After calculating the reliability values for all candidate neighbors, the node i selects a set of parents that satisfy the constraint (nodes with reliability values greater than or equal to the threshold are selected. The system manager determines the threshold). The best reliable node is selected as the parent—the node i selects the node with the lowest rank as the parent among candidates with the same reliability.

**Step 5 (Calculating rank of each node):** The rank of each node of the graph should decrease uniformly while moving up towards the BR, and it should increase uniformly while moving down towards the leaf nodes. Also, it should be limited to MIN-H[4] and MAX-H[5] [2]. Therefore, the BR sets its value to MAX-H to meet the rank uniformity. Each node calculates its N ranks R(N) using the total parent rank R(P) selected in step 4, and the reliability of the selected parent is calculated using Eq. (5).

$$R(N) = R(P) + (R_j(Final) \times 100) + MIN - H \tag{5}$$

---

[4] Min Hop Rank increase
[5] Ma Hop Rank Increase

**Step 6:** When a node calculates its reliability value, it selects its parent and calculates its rank. It updates the records, including its measures (reliability values of itself and its neighbors), and transmits its DIO to the neighboring nodes.

**Step 7:** Routing and transmitting the routing table to the neighboring nodes via the DIO message upon receiving the message.

An example of a DODAG graph is given in Figure 4.

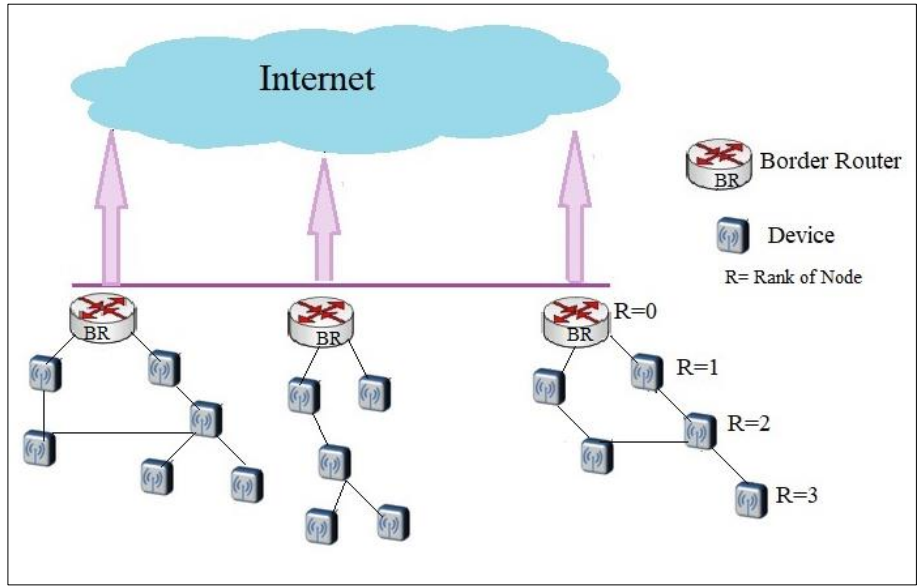

**FIGURE 4** An example of a DODAG graph created by the proposed method.

## 3.3 | Phase 2: Detecting Sinkhole in DSH-RPL

In the proposed DSH-RPL method, it is assumed that an IoT network has no malicious nodes while being established. A DODAG graph is created in the first phase using trustworthy nodes with high reliability. As the network continues operation and new nodes are added to the graph, an attack might occur. In the proposed method, the presence of a sinkhole node is checked before data transmission.

### 3.3.1 | Detecting Malicious Nodes based on Rank

In this method, the correct routing table of each node is broadcast in the established network before the sinkhole attacks. In the first step, two characteristics are defined for detecting abnormal DIO messages:

**Differences in node rank with parent (DNR-P):** this characteristic is the difference in the rank of the node with its parent, which is calculated according to Eq. (6). The value of this characteristic is obtained when the routing table is constructed or updated.

$$DNR-P = |(parent\_rank) - (node\_rank)| \qquad (6)$$

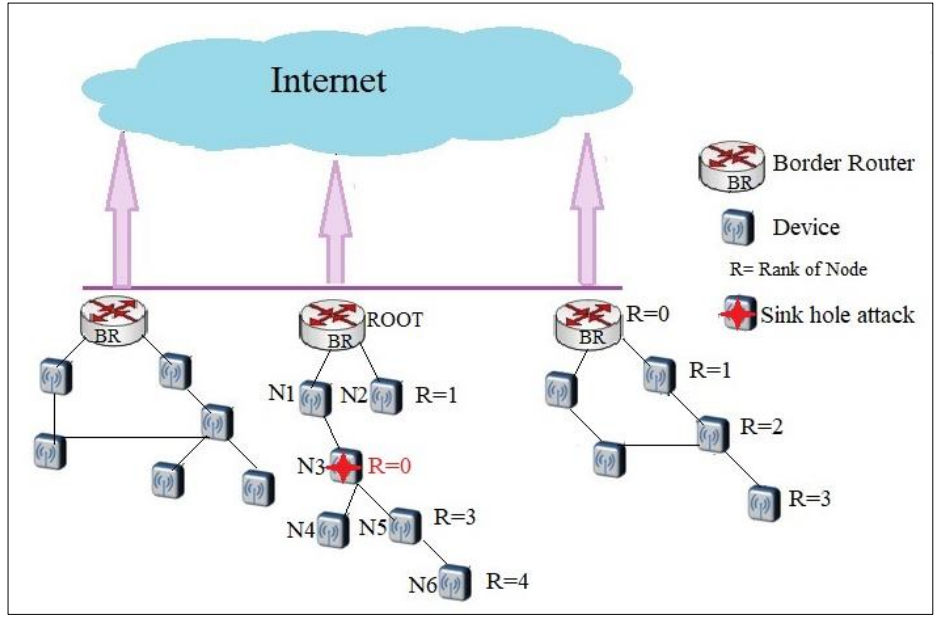

**FIGURE 5** The presence of sinkhole attack in the network.

For example, according to Figure (5), DNR-P of node N5 is 1, because its rank in the graph is 3. While its parent rank, N3, is 2. Therefore, DNR-P of N5 is $|2-3|=1$.

**The next characteristic is the difference between the source node and the node itself (DSN-NI):** this characteristic is the difference between the source node transmitting the message and the node itself as in Eq. (7).

$$DSN-NI = |(\text{Message sender node rank}) - (node\_rank)| \qquad (7)$$

For example, in Figure (5), N5 receives the message from N3, and N5 calculates DNS-NI. Since the rank of the node transmitting the DIO message (N3) is 2, DSN-NI would be 1. The rank of the transmitter node is 2, and the rank of the node itself is 3, $|2-3|=1$.

However, in Figure (5), when the sinkhole node penetrates the network, it announces its rank zero for DIO messages to introduce itself as the root node and transmit it to all children nodes. When N5 receives a DIO message from the malicious node (N3), it should calculate DSN-NI, which would be 3. Because the rank of the source node is zero, and the rank of N5 is 3. Therefore, DSN-NI in Figure (5) is $|0-3|=3$.

The DSH-RPL method considers the DIO message as a malicious node if DSN-NI>DNR-P. In Figure (5), the DIO message transmitted by the malicious node (N3) is detected as a malicious message by N5 using DSN-NI>DNR-P and informs the root node via a control message.

The proposed method applies another operation on the route, including a node suspicious as a sinkhole node to ensure that a node is malicious.

### 3.3.2 | Detecting the Malicious Node based on PDR

In the second step, the packet delivery rate (PDR) is used to ensure the detection of the sinkhole node. PDR represents the ratio of the packets delivered successfully to the number of packets transmitted by the transmitter node. A transmitter node confirms that a packet is transmitted successfully when it receives an ACK packet from the receiver node. To this end, after receiving the warning message, the root node transmits several control messages called RPL-MC to the leaf nodes of the route with suspicious nodes. This message includes the destination address via the route of interest (Base). Figure 6 shows the Format of RPL-MC message.

| 1B | 1B | 2B |
|---|---|---|
| Type | Code | Checksum |
| Base ||| 
| Option |||

**FIGURE 6** Format of RPL-MC message

Upon receiving RPL-MC, the nodes should transmit an ACK message to the root node. When a sinkhole node exists along the route, all or some of the messages are discarded and not transmitted to the destination by receiving the RPL-MC message. This can be understood using PDR.

PDR can be calculated by comparing the number of RPL-MC packets with the number of received ACK packets. Therefore, the average PDR is calculating using Eq. (8).

$$PDR = \left( \frac{\sum \text{Number of ACKs received from the destination}}{\sum \text{Number of RPL-MC packets sent}} \right) \qquad (8)$$

In the proposed method, a statistical method is used to determine abnormal conditions and the presence of sinkhole node in the network through defining a threshold value (PDRT). In this method, a lower threshold limit and an upper threshold limit called LT-P and UT-P, respectively, are considered, and the average PDR is calculated using Eq. (8). PDRA is calculated, and a normal distribution of the threshold is obtained using Eq. (9) by calculating the standard deviation.

$$PDR_T = PDR_A - SD \qquad (9)$$

The values that exceed the interval between LT-P and UT-P are determined as abnormal values. Considering Eq. (9), if the PDR value of the route with a suspicious node is lower than PDRT, PDR is lower than the expected value based on the received ACKs. Thus, there exists a sinkhole node along the route, which is not allowed to transmit packets to the other nodes and these packets are discarded by the sinkhole node. Therefore, the malicious node should be omitted from the routing operation. This problem is studied in the third phase.

## 3.4 | Phase 3: Quarantining the Malicious Node in DSH-RPL

In this phase, the sinkhole node is quarantined after detection and isolated from other nodes, so that routing and data transmission is not interrupted. To this end, the root node, which is aware of the sinkhole attack and has received the number of the sinkhole node, generates a warning message and transmits it to all nodes of its graph. Also, this node quarantines the adversary node. The primary information broadcast by the message transmitted from the root node to the other nodes, includes the rank of the malicious node that allows other nodes to reorganize themselves. Therefore, the malicious node is isolated from the network, and the connected nodes are reorganized to create a new graph, in which there is no sinkhole node.

## 3.5 | Phase 4: Data Transmission based on Encryption

Encryption is used to prevent illegal access, and ensure secure information transmission. Encryption is the process of converting information or data to code. The new security factor and the functions used to simulate the performance of the encryption strategies are discussed in this section. The Homomorphic Encryption (HE) technique is used to execute data security. Encrypting means that an encryption hash function is executed using an encryption algorithm. The encrypted data is transmitted to the final device, and the data is decoded, and their reliability is validated.

### 3.5.1 | Employing HE

HE is an encryption method that allows various specific computations to be applied to the encrypted data and obtain an encrypted text. The receiver might decode the encrypted text and obtain the aggregated data without knowing the primary data. Therefore, this method is suitable for transmitting objects to the root in DODAG. This encryption method provides the possibility of applying computations to the encrypted data. The HE processes include 4 functions:

**Key generation:** the transmitter generates a public key ($P_k$) and a secret key ($S_k$) for encrypting the primary text.

**Encryption:** using the secret key ($S_k$), the transmitter encrypts the simple text ($T_s$) and creates s $S_k(T_s)$. The encrypted text ($T_e$) and the public key are transmitted to the receiver.

**Evaluation:** The receiver has a function f to evaluate the encrypted text, and this function is applied using $P_k$ to perform the decoding operation.

**Decoding:** The text TE received by the receiver is decoded using the secret key ($S_k$), and the result is obtained.

Therefore, the data transmission from the objects to the root and vice versa is performed by HE encryption in the routes without malicious nodes. HE process is an encryption strategy in which the process can be done by the users themselves. The primary message is restored using the secret key, and the messages are converted in the nodes using public and secret keys. The flowchart of the proposed DSH-RPL method is shown in Figure 7.

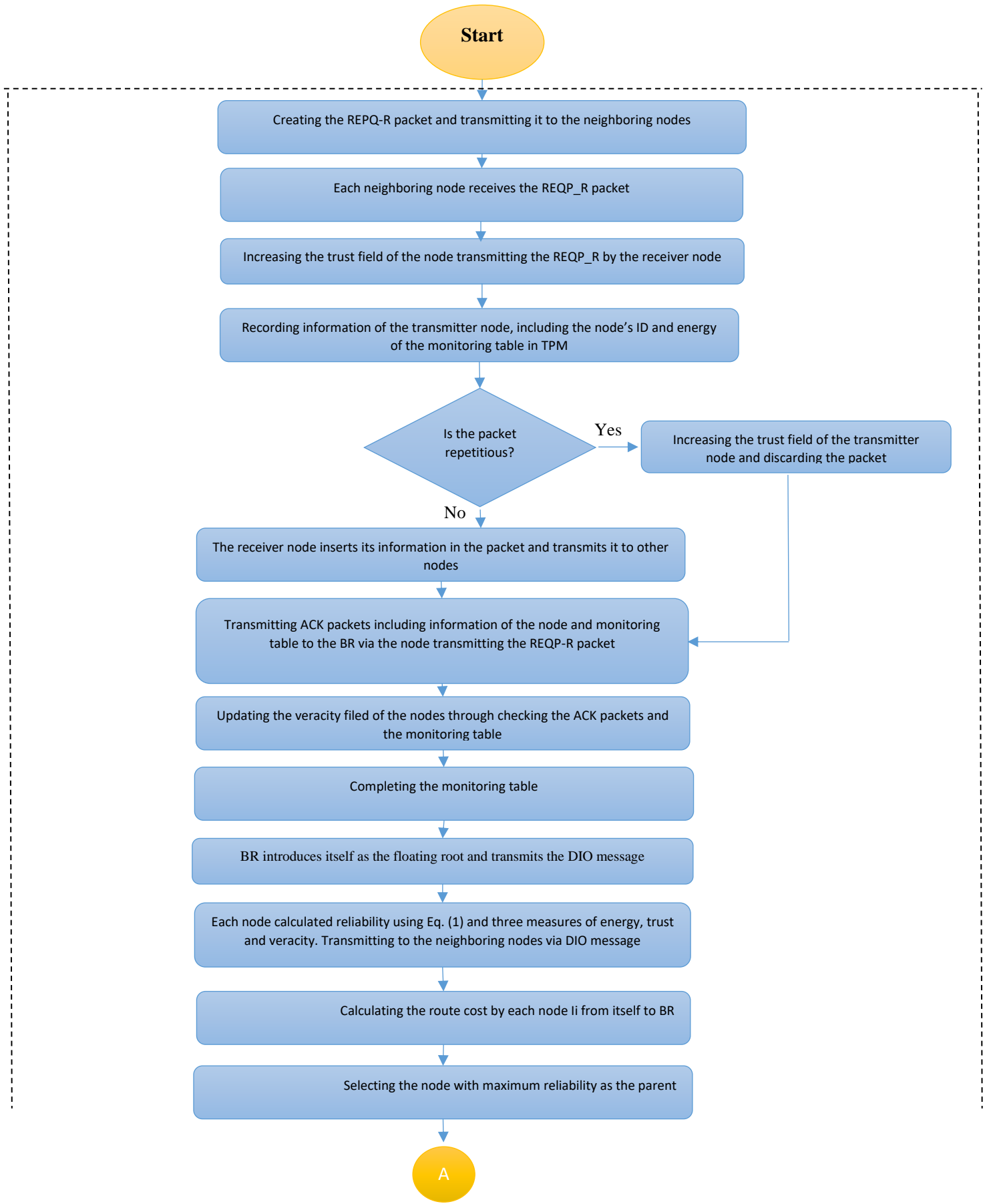

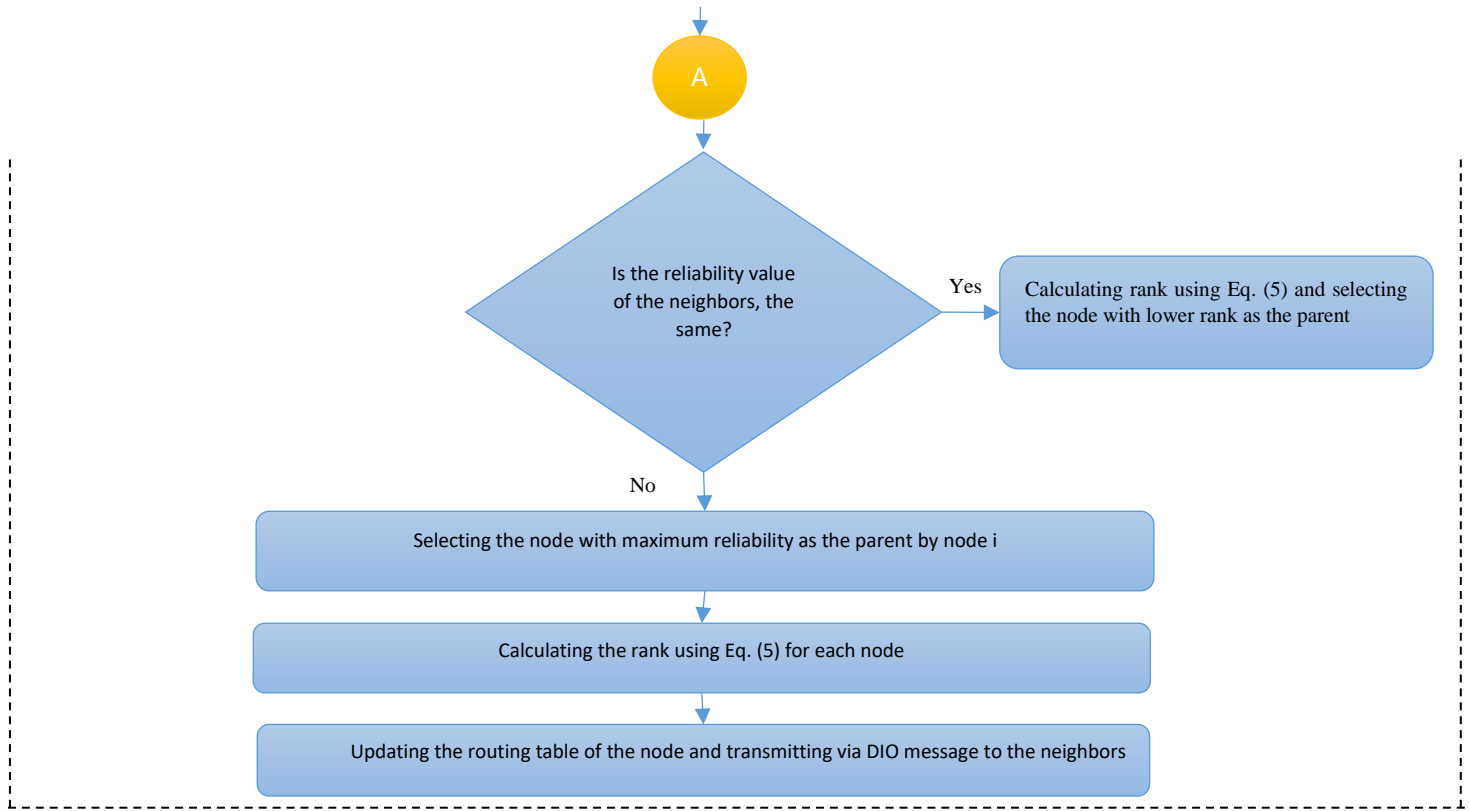
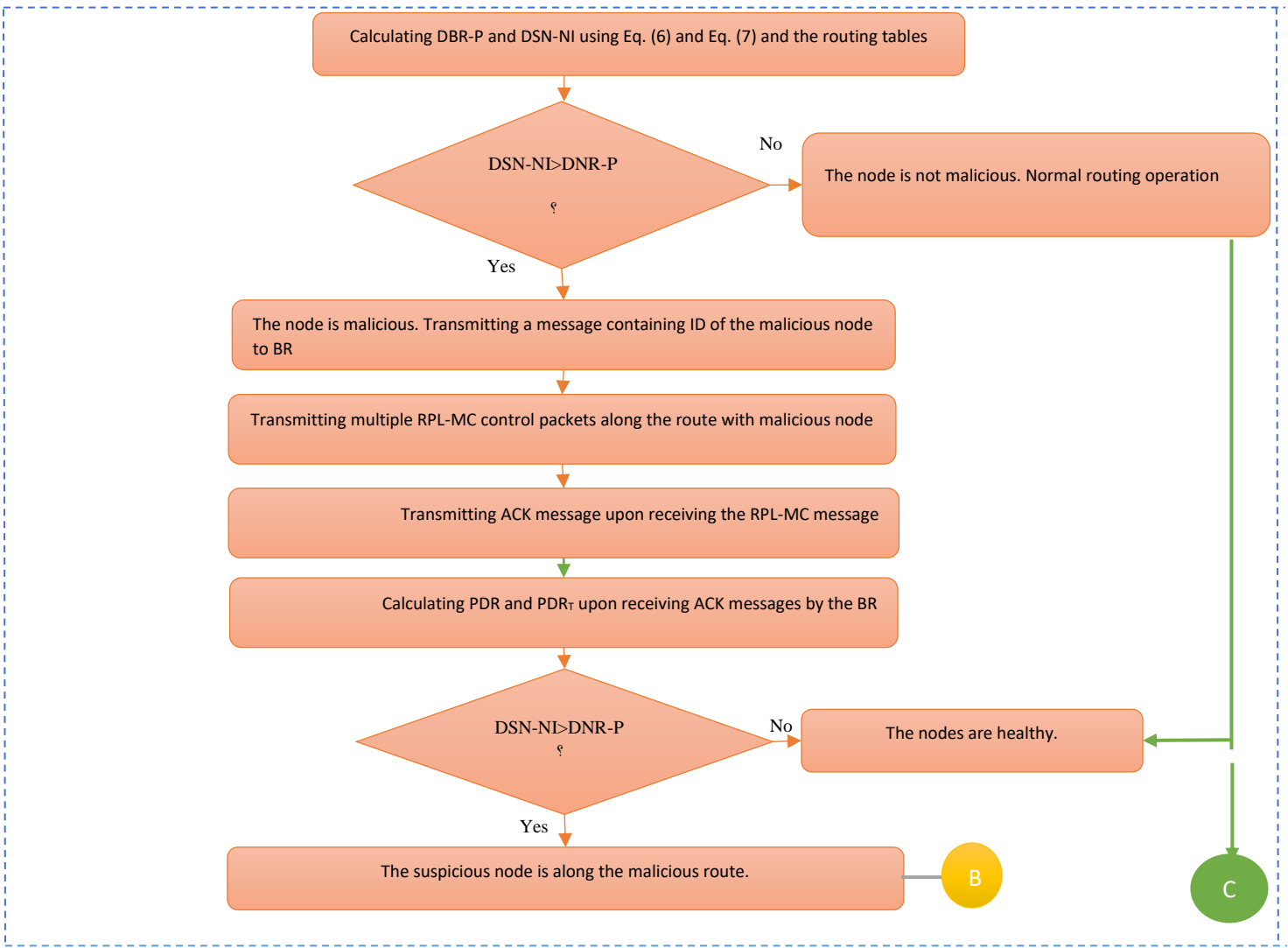

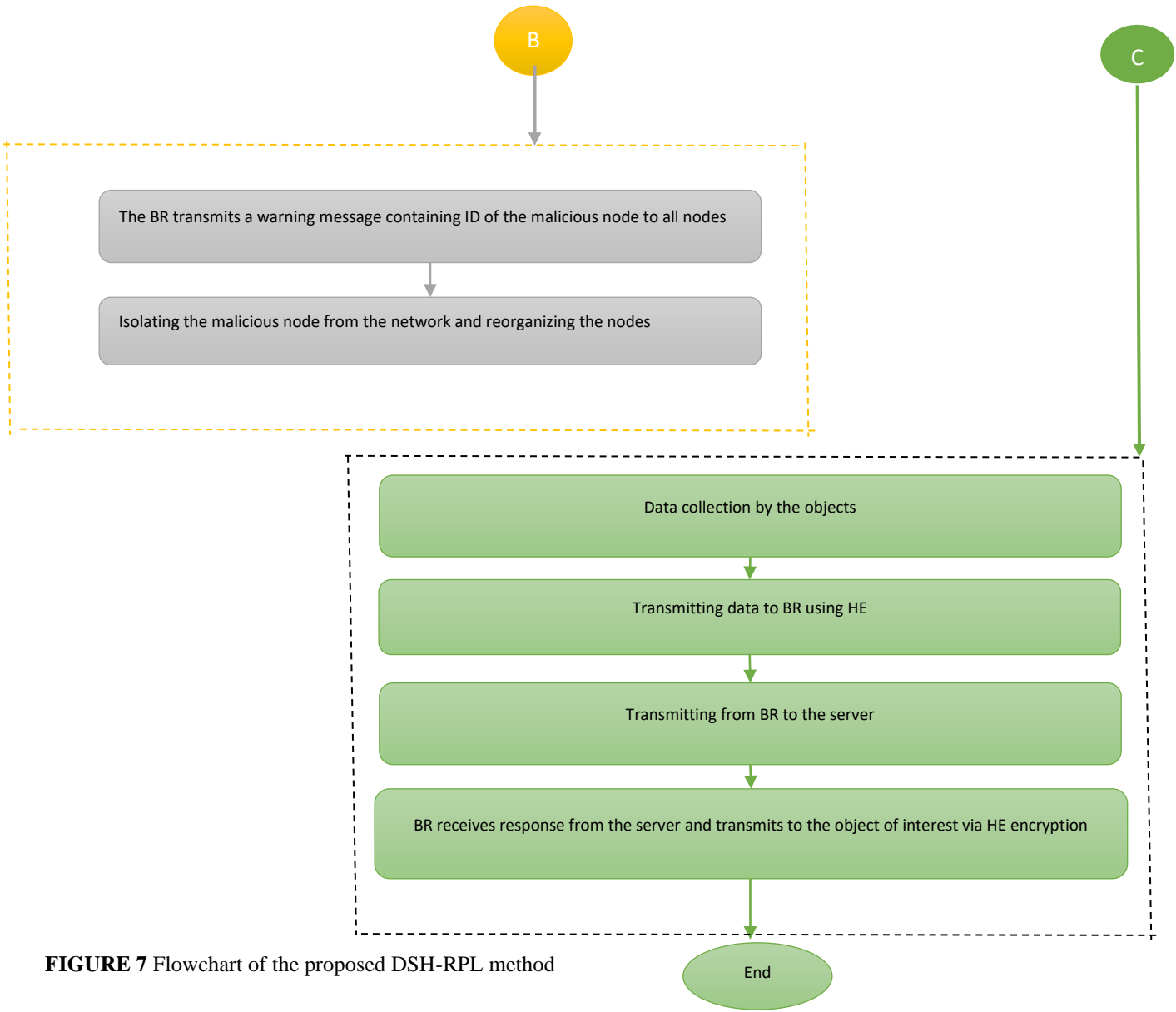

**FIGURE 7** Flowchart of the proposed DSH-RPL method

## 4 | PERFORMANCE EVALUATION

This section analyzes security parameters and evaluates the qualitative performance in the form of numerical results to validate the performance of the proposed DSH-RPL method. To demonstrate a feasibility study, the performance analysis of DSH-RPL has been divided into three parts:

1) Detection Rate (DR),
2) False Negative Rate (FNR),
3) False Positive Rate (FPR),
4) Packet Delivery Rate (PDR)

### 4.1 | PERFORMANCE METRICS

The proposed DSH-RPL method has been simulated and its performance evaluated in Network Simulator version 2 (NS-2) running on Linux Ubuntu 14.03 LTS. The results were compared with both methods (SecTrust-RPL [1] and IBOOS-RPL [3]).

DR: Ratio of sinkhole nodes to total malicious nodes that were correctly diagnosed as a sinkhole attack. Eq. (10) determines the DR [14-17].

$$DR = \left(\frac{TPR}{TPR+FNR}\right)*100 \quad \text{where} \quad All = TPR+TNR+FPR+FNR \tag{10}$$

FPR: The FPR is determined by the total number of nodes mistakenly found as the malevolent nodes divided by the total number of usual nodes [18-20]. Hence, Eq. (11) illustrates the

$$FPR = \left(\frac{FPR}{FPR+TNR}\right)*100 \quad \text{Where:} \quad TNR = \left(\frac{TNR}{TNR+FPR}\right)*100 \tag{11}$$

FNR: The rate of the malevolent node to total normal nodes incorrectly signed as a normal node [21-25]. The calculation is proved by Eq. (12).

$$FNR = \left(\frac{TPR+TNR}{All}\right)*100 \quad \text{Where:} \quad TPR = \left(\frac{TPR}{TPR+FNR}\right)*100 \tag{12}$$

PDR: This criterion represents the rate of packets that were successfully delivered to the destination [26-29]. Eq. (13) determines the PDR.

$$PDR = \frac{1}{n}*\frac{\sum_{i=1}^{n} X_i}{\sum_{i=1}^{n} Y_i}*100\% \tag{13}$$

### 4.2 | SIMULATION RESULT

All three methods are evaluated according to Table 1 under four scenarios. Table 2 displays the significant parameters used in the simulation.

**TABLE 1:** Parameters used.

| Parameters | Value |
| --- | --- |
| Operating System | Linux Ubuntu |
| IP version | IPV6 |
| Protocol | RPL |
| Simulation time | 1000 second |
| Number of nodes | 500 |
| Type of attacks | Sinkhole |
| Average transaction size | 77 Byte |

A comparison of all three methods under four criteria versus the Attack interval is shown in Tables 3 to 6.

**TABLE 2** Parameters used for four scenarios.

| Scenario #1 | | Scenario #2 | |
|---|---|---|---|
| Sinkhole rate | 10% | Sinkhole rate | 20% |
| Topology (m x m) | 300*300 | Topology (m x m) | 300*300 |
| Time | 1000 | Time | 1000 |
| Scenario #3 | | Scenario #4 | |
| Sinkhole rate | 30% | Attack interval | 0.5, 1, 1.5, 2, 2.5, 3, 3.5 |
| Topology (m x m) | 300*300 | Topology (m x m) | 300*300 |
| Time | 1000 | Time | 1000 |

**TABLE 3** DR vs Attack interval (30% attack)

| Attack interval | DR (%) | | |
|---|---|---|---|
| | IBOOS-RPL | SecTrust-RPL | DSH-RPL |
| 0.5 | 62 | 70 | 86 |
| 1 | 63 | 71 | 88 |
| 1.5 | 66 | 72 | 91 |
| 2 | 69 | 74 | 92 |
| 2.5 | 72 | 77 | 93 |
| 3 | 72 | 78 | 94 |
| 3.5 | 74 | 81 | 96 |

**TABLE 4** FNR vs Attack interval (30% attack)

| Attack interval | FNR (%) | | |
|---|---|---|---|
| | IBOOS-RPL | SecTrust-RPL | DSH-RPL |
| 0.5 | 15.5 | 12.4 | 7.2 |
| 1 | 17.8 | 13.3 | 7.65 |
| 1.5 | 17.3 | 14.8 | 7.8 |
| 2 | 18.67 | 15.4 | 8.5 |
| 2.5 | 18.8 | 15.7 | 9.5 |
| 3 | 19.4 | 14.7 | 9.8 |
| 3.5 | 21.6 | 15.6 | 10.98 |

**TABLE 5** FPR vs Attack interval (30% attack)

| Attack interval | FPR (%) | | |
|---|---|---|---|
| | IBOOS-RPL | SecTrust-RPL | DSH-RPL |
| 0.5 | 23.6 | 18.5 | 10.4 |
| 1 | 24.7 | 19.8 | 11.3 |
| 1.5 | 24.5 | 20.9 | 12.8 |
| 2 | 25.1 | 20.67 | 12.4 |
| 2.5 | 25.3 | 20.8 | 12.7 |
| 3 | 26.5 | 21.4 | 13.2 |
| 3.5 | 27.1 | 23.6 | 13.6 |

**TABLE 6** PDR vs Attack interval (30% attack)

| Attack interval | PDR (%) | | |
|---|---|---|---|
| | IBOOS-RPL | SecTrust-RPL | DSH-RPL |
| 0.5 | 66 | 74 | 82 |
| 1 | 67 | 75 | 84 |
| 1.5 | 68 | 76 | 87 |
| 2 | 69 | 77 | 89 |
| 2.5 | 72 | 79 | 90 |
| 3 | 73 | 80 | 93 |
| 3.5 | 75 | 82 | 98 |

**DR**: Figure 8 shows the DR in all three scenarios. As can be seen in diagrams A to C, in all simulated scenarios, the proposed DSH-RPL method has a higher detection rate compared to IBOOS-RPL, and SecTrust-RPL, because in the proposed method, the malicious sinkhole nodes are detected in two steps using precise evaluations. In most of the previous methods like IBOOS-RPL, and SecTrust-RPL, only one examination is made to detect the malicious sinkhole node, and the node is considered as malicious only with a straightforward validation step. However, in the proposed method, the nodes are first examined in terms of their rank, such that the difference between the rank of the transmitter node and the receiver node is examined; if $P\ DSN - NI$, the node is considered as malicious. Second, another examination is made in terms of packet delivery rate for more precise detection of the malicious node. In other words, it examines the number of transmitted control packets to detect the sinkhole node precisely. These factors increase the DR of the proposed DSH-RPL method compared to SecTrust-RPL and IBOOS-RPL by 26% and 31%, respectively, and it is higher than the other two methods in all three diagrams.

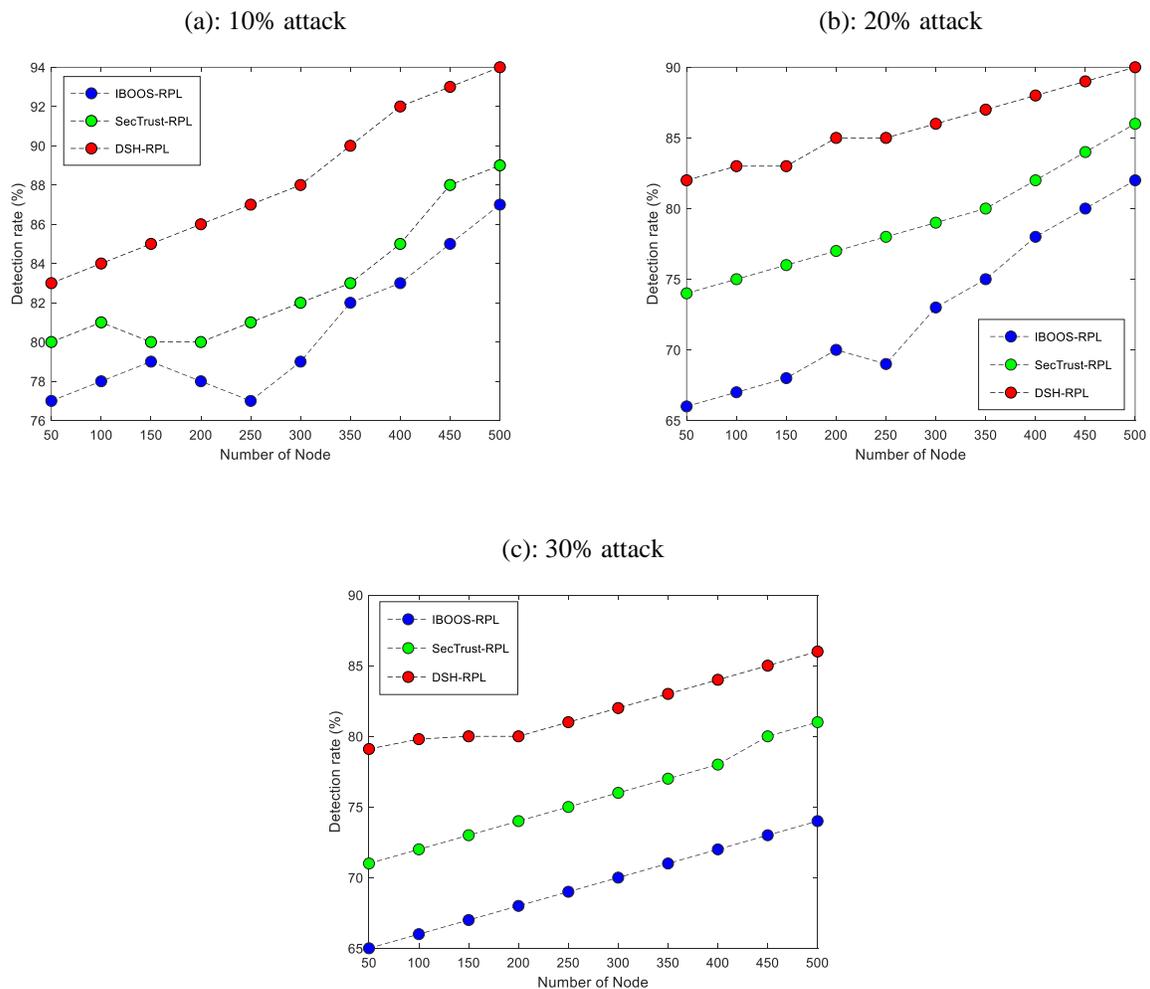

**FIGURE 8** Number of Nodes vs DR.

**FNR**: The results of the FNR criterion are given in all three scenarios as shown in Figure 9. This measure represents the number of nodes that are not malicious, but they are incorrectly detected as malicious nodes. According to the information obtained from simulation, as can be seen in diagrams A to C, FNR of the proposed method is less than IBOOS-RPL and SecTrust-RPL, because the proposed DSH-RPL performs a complete two-step examination to detect the malicious sinkhole nodes, and the decision is not made only based on one factor. In the SecTrust method, the trust between the direct and the indirect nodes is examined to detect the malicious node. In the proposed DSH-RPL method, a precise two-step examination is made to detect the malicious sinkhole nodes; in the first step, the malicious node is detected based on the nodes' rank. In the second step, the control packets are transmitted to the suspicious nodes, and the average PDR is calculated upon receiving an ACK. PDR is used to detect and quarantine the sinkhole node. As a result of this two-step investigation, the proposed DSH-RPL has less FNR compared to the other two methods, such that it is improved by 16.1% and 22.78% compared to the SecTrust-RPL and IBOOS-RPL respectively.

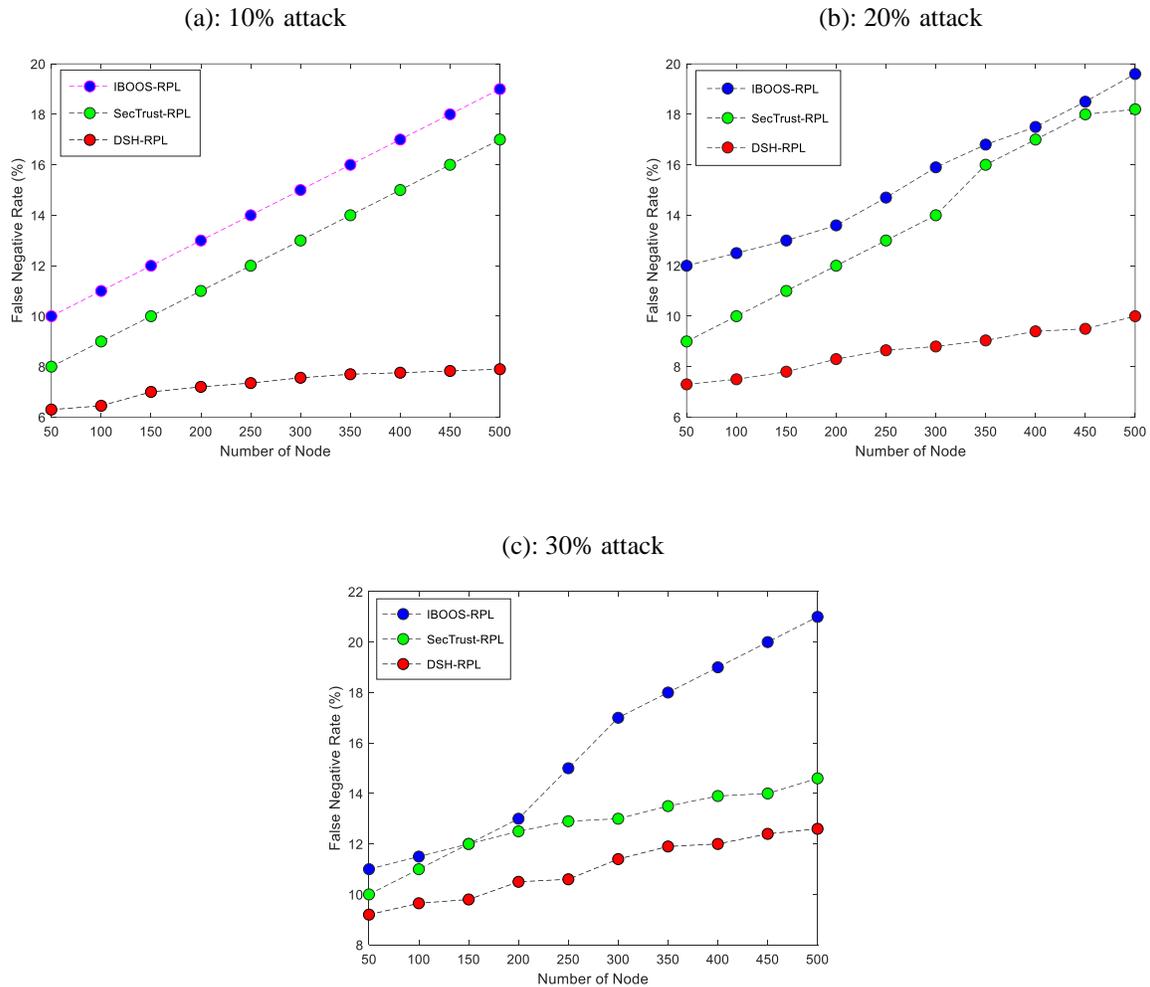

**FIGURE 9** Number of Nodes vs FNR.

**FPR**: Figure 10 shows the FPR in all three scenarios. This measure represents the percentage of the malicious nodes that are not detected. According to diagrams A to C, the proposed DSH-RPL has less FPR than the other two methods. FPR of the IBOOS-RPL is higher than the other two methods, because it makes a superficial comparison to detect malicious activities. Therefore, many of the malicious nodes are not detected correctly. However, in the proposed DSH-RPL method, the malicious node is detected in two steps. First, the nodes that have not transmitted their rank correctly are detected as malicious nodes using the information of the node recorded while creating a reliable DODAG in each node. However, for more reliability and more precise detection of the malicious node, the proposed method examines the PDR along the routes with suspicious nodes to precisely detect the malicious nodes. These two steps, along with each other and the information inserted in each node while generating a reliable DODAG, reduces the FPR of the proposed DSH-RPL compared to the other two methods in all scenarios. FPR of the DSH-RPL is improved by 18.2% and 23.1% compared to SecTrust-RPL and IBOOS-RPL.

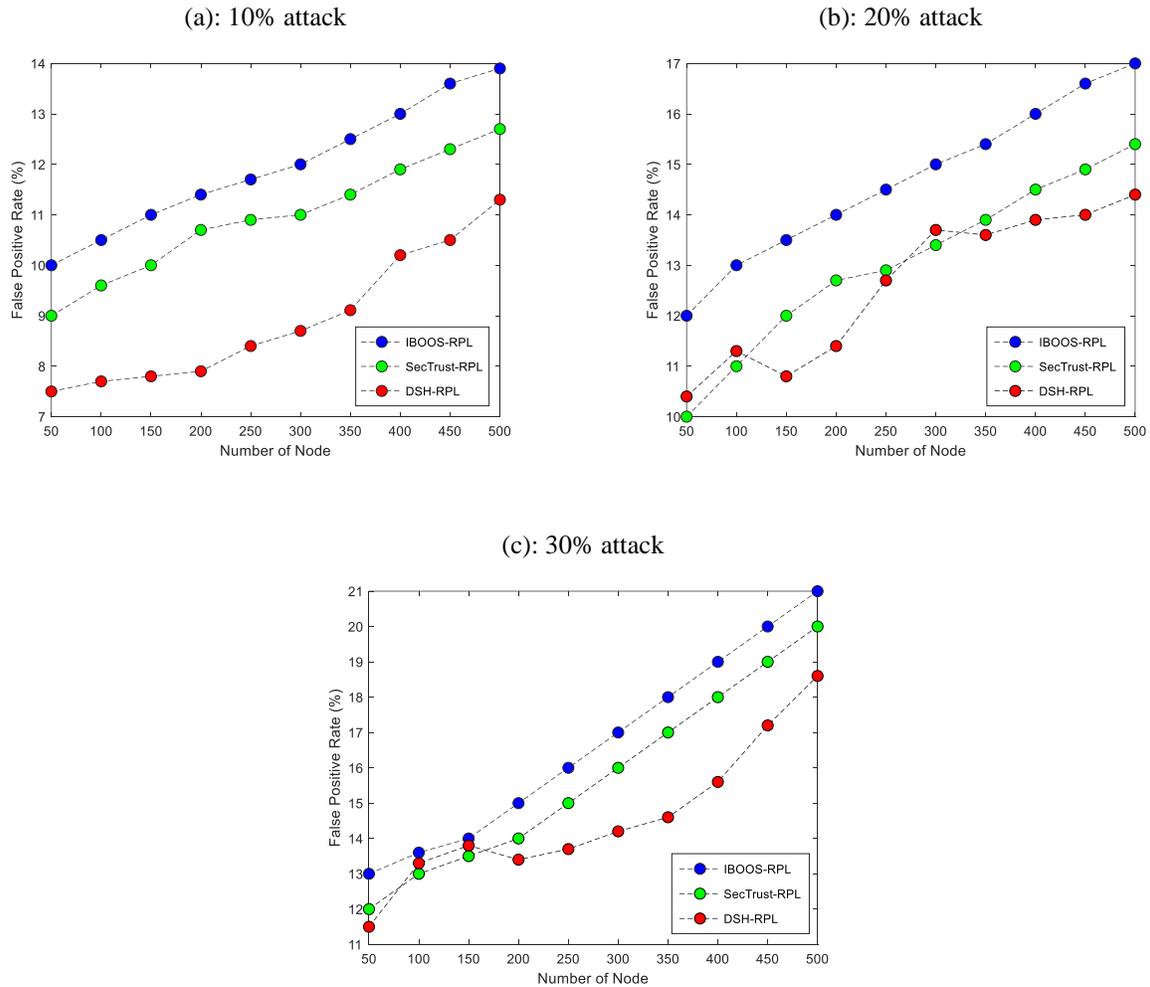

**FIGURE 10** Number of Nodes vs FPR.

**PDR:** The results of the PDR criterion are given in all three scenarios as shown in Figure 11. As can be seen in the diagrams of different scenarios, on average, the proposed DSH-RPL is improved by 19.68% and 25.32% compared to SecTrust-RPL and IBOOS-RPL, respectively. The PDR improvement in the proposed method is because it employs a reliable node to generate DODAG such that the reliability of the nodes is examined, and the nodes are evaluated based on their reliability in packet transmission. Also, the proposed method detects the malicious sinkhole nodes in two steps and isolates them from the network to avoid interrupting the data transmission. The proposed DSH-RPL also encrypts data to increase the reliability of data transmission. The proposed method transmits data using homomorphic encryption; thus, if a malicious node aims to destruct data, it cannot succeed. All these factors increase the PDR of the proposed DSH-RPL method compared to the other two methods.

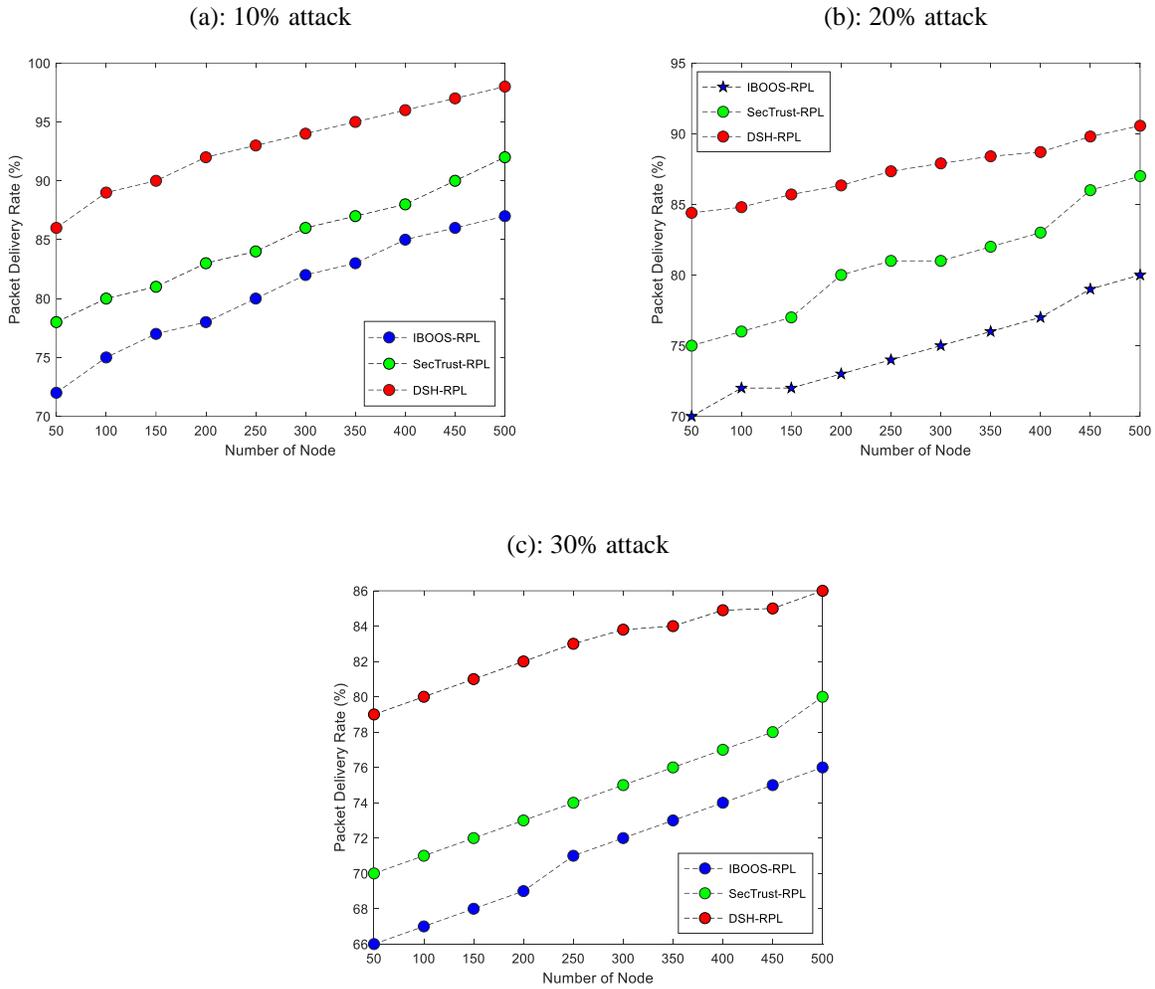

**FIGURE 11** Number of Nodes vs PDR.

## 5 | CONCLUSION

The IoT ecosystem operates in an entirely distributed environment with a low delay so that the devices can communicate with each other securely and exchange time-sensitive data. In this paper, homomorphic encryption is used to encrypt data; thus, the data is transmitted from objects to the root and vice versa using HE encryption along the routes free of malicious nodes. HE encryption is an encryption strategy in which processing can be carried out by the users. The main message is restored using the private key, and the messages are transformed into the nodes using the public and private keys. The proposed DSH-RPL is implemented in several IoT scenarios. The security requirements and an attack model are defined to evaluate the efficiency of the DSH-RPL and its capability to realize these requirements. The simulation results show that the DSH-RPL improves the false-positive, false-negative, packet-delivery rate, and detection rate, significantly. This capability of DSH-RPL provides the possibility of secure data transmission among IoT devices.

## CONFLICT OF INTEREST

None.

# DATA Availability Statement

The data of this paper is the result of simulation and all the data are presented in the form of graphs inside the paper. There is no private data in this article.

# REFERENCE

1. Airehrour, D., Gutierrez, J. A., & Ray, S. K. (2019). SecTrust-RPL: A secure trust-aware RPL routing protocol for Internet of Things. Future Generation Computer Systems, 93, 860-876.
2. Thigale, S. B., Pandey, R., & Dhotre, V. A. (2020). Robust Routing for Secure Communication in Internet of Things Enabled Networks. In Techno-Societal 2018 (pp. 79-86). Springer, Cham.
3. Nikravan, M., Movaghar, A., & Hosseinzadeh, M. (2018). A lightweight defense approach to mitigate version number and rank attacks in low-power and lossy networks. Wireless Personal Communications, 99(2), 1035-1059.
4. Bhale, P., Dey, S., Biswas, S., & Nandi, S. (2020, January). Energy Efficient Approach to Detect Sinkhole Attack Using Roving IDS in 6LoWPAN Network. In International Conference on Innovations for Community Services (pp. 187-207). Springer, Cham.
5. Dorri, A., Kanhere, S. S., Jurdak, R., & Gauravaram, P. (2019). LSB: A Lightweight Scalable Blockchain for IoT security and anonymity. Journal of Parallel and Distributed Computing, 134, 180-197.
6. Tahir, S., Bakhsh, S. T., & Alsemmeari, R. A. (2019). An intrusion detection system for the prevention of an active sinkhole routing attack in Internet of things. International Journal of Distributed Sensor Networks, 15(11), 1550147719889901.
7. Khalid, U., Asim, M., Baker, T., Hung, P. C., Tariq, M. A., & Rafferty, L. (2020). A decentralized lightweight blockchain-based authentication mechanism for IoT systems. Cluster Computing, 1-21.
8. Alshehri, M. D., & Hussain, F. K. (2019). A fuzzy security protocol for trust management in the internet of things (Fuzzy-IoT). Computing, 101(7), 791-818.
9. Christidis, K., & Devetsikiotis, M. (2016). Blockchains and smart contracts for the internet of things. Ieee Access, 4, 2292-2303.
10. Ravi, N., & Shalinie, S. M. (2020). Learning-driven detection and mitigation of DDoS attack in IoT via SDN-cloud architecture. IEEE Internet of Things Journal, 7(4), 3559-3570.
11. Ouaddah, A., Abou Elkalam, A., & Ait Ouahman, A. (2016). FairAccess: a new Blockchain-based access control framework for the Internet of Things. Security and Communication Networks, 9(18), 5943-5964.
12. Murali, S., & Jamalipour, A. (2019). A lightweight intrusion detection for sybil attack under mobile RPL in the internet of things. IEEE Internet of Things Journal, 7(1), 379-388.
13. Zaminkar, M., & Fotohi, R. (2020). SoS-RPL: Securing Internet of Things Against Sinkhole Attack Using RPL Protocol-Based Node Rating and Ranking Mechanism. WIRELESS PERSONAL COMMUNICATIONS.
14. Mabodi, K., Yusefi, M., Zandiyan, S., Irankhah, L., & Fotohi, R. (2020). Multi-level trust-based intelligence schema for securing of internet of things (IoT) against security threats using cryptographic authentication. The Journal of Supercomputing, 1-26.
15. Raoof, A., Matrawy, A., & Lung, C. H. (2018). Routing attacks and mitigation methods for RPL-based Internet of Things. IEEE Communications Surveys & Tutorials, 21(2), 1582-1606.
16. Ande, R., Adebisi, B., Hammoudeh, M., & Saleem, J. (2020). Internet of Things: Evolution and technologies from a security perspective. Sustainable Cities and Society, 54, 101728.
17. Seyedi, B., & Fotohi, R. (2020). NIASHPT: a novel intelligent agent-based strategy using hello packet table (HPT) function for trust Internet of Things. The Journal of Supercomputing, 1-24.
18. Parra, G. D. L. T., Rad, P., Choo, K. K. R., & Beebe, N. (2020). Detecting Internet of Things attacks using distributed deep learning. Journal of Network and Computer Applications, 102662.
19. Bhattacharjya, A., Zhong, X., Wang, J., & Li, X. (2020). CoAP—application layer connection-less lightweight protocol for the Internet of Things (IoT) and CoAP-IPSEC Security with DTLS Supporting CoAP. In Digital Twin Technologies and Smart Cities (pp. 151-175). Springer, Cham.

20. Fotohi, R., Firoozi Bari, S., & Yusefi, M. (2020). Securing wireless sensor networks against denial-of-sleep attacks using RSA cryptography algorithm and interlock protocol. International Journal of Communication Systems, 33(4), e4234.
21. Farzaneh, B., Montazeri, M. A., & Jamali, S. (2019, April). An anomaly-based IDS for detecting attacks in RPL-based internet of things. In 2019 5th International Conference on Web Research (ICWR) (pp. 61-66). IEEE.
22. Farzaneh, B., Koosha, M., Boochanpour, E., & Alizadeh, E. (2020, April). A New Method for Intrusion Detection on RPL Routing Protocol Using Fuzzy Logic. In 2020 6th International Conference on Web Research (ICWR) (pp. 245-250). IEEE.
23. Faraji-Biregani, M., & Fotohi, R. (2020). Secure communication between UAVs using a method based on smart agents in unmanned aerial vehicles. The Journal of Supercomputing, 1-28.
24. Farzaneh, B., Ahmed, A. K., & Alizadeh, E. (2019, October). MC-RPL: A New Routing Approach based on Multi-Criteria RPL for the Internet of Things. In 2019 9th International Conference on Computer and Knowledge Engineering (ICCKE) (pp. 420-425). IEEE.
25. Roldán, J., Boubeta-Puig, J., Martínez, J. L., & Ortiz, G. (2020). Integrating complex event processing and machine learning: An intelligent architecture for detecting IoT security attacks. Expert Systems with Applications, 149, 113251.
26. Fotohi, R., Nazemi, E., & Aliee, F. S. (2020). An Agent-Based Self-Protective Method to Secure Communication between UAVs in Unmanned Aerial Vehicle Networks. Vehicular Communications, 100267.
27. Qureshi, K. N., Rana, S. S., Ahmed, A., & Jeon, G. (2020). A novel and secure attacks detection framework for smart cities industrial internet of things. Sustainable Cities and Society, 61, 102343.
28. Fotohi, R. (2020). Securing of Unmanned Aerial Systems (UAS) against security threats using human immune system. Reliability Engineering & System Safety, 193, 106675.
29. Parra, G. D. L. T., Rad, P., Choo, K. K. R., & Beebe, N. (2020). Detecting Internet of Things attacks using distributed deep learning. Journal of Network and Computer Applications, 102662.
30. Zeadally, S., & Tsikerdekis, M. (2020). Securing Internet of Things (IoT) with machine learning. International Journal of Communication Systems, 33(1), e4169.
31. Odusami, M., Misra, S., Abayomi-Alli, O., Abayomi-Alli, A., & Fernandez-Sanz, L. (2020). A survey and meta-analysis of application-layer distributed denial-of-service attack. International Journal of Communication Systems, e4603.